\documentclass[prd,showpacs,showkeys,nofootinbib,floatfix,eqsecnum,
               fleqn,preprint,12pt,tightenlines]{revtex4-1}

\newcommand{\version}{v4}           

\usepackage{amsmath,amssymb,revsymb,graphicx,dcolumn}
\usepackage{array}
\usepackage{hyperref}

\newcommand{\beq}{\begin{equation}}
\newcommand{\eeq}{\end{equation}}
\newcommand{\beqa}{\begin{eqnarray}}
\newcommand{\eeqa}{\end{eqnarray}}
\newcommand{\bsubeqs}{\begin{subequations}}
\newcommand{\esubeqs}{\end{subequations}}

\begin{document}

\begin{widetext}
%
\noindent arXiv:2005.12157 \hfill KA--TP--03--2020\;(\version)
%
%
\newline\vspace*{5mm}
\end{widetext}

\title{Another model for the regularized big bang\vspace*{4mm}}

\author{F.R. Klinkhamer}
\email{frans.klinkhamer@kit.edu}

\affiliation{Institute for Theoretical Physics,
Karlsruhe Institute of Technology (KIT),\\
76128 Karlsruhe, Germany\\}

\vspace*{5mm}

\begin{abstract}
\vspace*{1mm}\noindent
We propose a gravitational model with a Brans--Dicke-type
scalar field having, in the would-be action,
a ``wrong-sign'' kinetic term
and a quartic interaction term. In a cosmological context,
we obtain, depending on the boundary conditions, either the
Friedmann solution or a kink-bounce solution.
The expanding-universe Friedmann solution has a
big bang curvature singularity, whereas the kink-bounce solution 
has a nonsingular bouncing behavior of the cosmic scale factor.
The bounce occurs precisely
at the moment when the scalar field of the kink-type configuration
goes through zero, making for a vanishing effective gravitational coupling.
\vspace*{10mm}
\end{abstract}



\maketitle

\section{Introduction}
\label{sec:Introduction}

Recently, we have shown that the singular Friedmann
\emph{solution}~\cite{Friedmann1922-1924,Weinberg1972,%
MisnerThorneWheeler2017,Wald1984}
from Einstein's general theory of relativity
can be modified (regularized) by considering a degenerate
metric with a nonzero length scale $b$~\cite{Klinkhamer2019,Klinkhamer2019-More}
(cosmological aspects have been studied in
Refs.~\cite{KlinkhamerWang2019-PRD,KlinkhamerWang2020-PRD}).

The goal, here, is to look for a modified version
of the \emph{theory}, which gives similar results.
This turns out to be quite 
difficult,
but perhaps this was to be expected, as we are trying to model
an entirely new phase from which classical spacetime and the universe
are supposed to emerge
(see also the discussion in App.~B of Ref.~\cite{Klinkhamer2019-More},
which contains further references).

Throughout, we use the metric signature $(-,+,+,+)$,
curvature conventions from Ref.~\cite{Weinberg1972},
and natural units with $c=\hbar=1$.

\section{Motivation}
\label{sec:Motivation}

The action of Einstein's general theory of relativity reads as
follows~\cite{Weinberg1972,MisnerThorneWheeler2017,Wald1984}:
\bsubeqs\label{eq:SofGR-calLofGR}
\beqa\label{eq:SofGR}
S &=& \int d^4 x \,\sqrt{-g}  \,\mathcal{L}\,,
\\[2mm]
\label{eq:calLofGR}
\mathcal{L} &=& -\frac{1}{16 \pi G_N}\,R+ \mathcal{L}_{M}\,,
\eeqa
\esubeqs
where $g$ is the determinant of the metric $g_{\mu\nu}$,
$R$ the Ricci curvature scalar,
$G_N>0$ Newton's gravitational coupling constant, and
$\mathcal{L}_{M}$ the Lagrange density of the standard matter
(the standard-matter fields are generically denoted by $\psi$).

In Ref.~\cite{Klinkhamer2019}, we proposed the following
degenerate-metric \textit{Ansatz} for a spatially flat universe:%
\bsubeqs\label{eq:mod-RW}
\beqa\label{eq:mod-RW-ds2}
\hspace*{-0mm}
ds^{2}
&\equiv&
g_{\mu\nu}(x)\, dx^\mu\,dx^\nu
=
- \frac{ t^{2}}{\widehat{b}^{\,2}+ t^{2}}\,d t^{2}
+ a^{2}(t)
\;\delta_{mn}\,dx^{m}\,dx^n\,,
\\[2mm]
\hspace*{-0mm}
\widehat{b} &>& 0\,,
\\[2mm]
\label{eq:mod-RW-cosmic-scale-factor}
\hspace*{-0mm}
a(t) &\in& \mathbb{R}\,,
\\[2mm]
\label{eq:mod-RW-range-cosmic-time-coordinate}
\hspace*{-0mm}
 t  &\in& (-\infty,\,\infty)\,,
\\[2mm]
\hspace*{-0mm}
x^{m} &\in& (-\infty,\,\infty)\,,
\eeqa
\esubeqs
where the spatial indices $m$, $n$
run over $\{1,\, 2,\, 3 \}$
and the \textit{Ansatz} parameter $\widehat{b}$
now carries a hat, in order to distinguish 
it from the model parameter $b$ introduced later on.
The metric from \eqref{eq:mod-RW} is degenerate,
having a vanishing determinant at $ t=0$, and describes
a spacetime defect with characteristic length scale $\widehat{b}\,$;
see Ref.~\cite{Klinkhamer2019-JPCS} for a review
of this type of spacetime defect.

The standard Einstein gravitational field
equation from \eqref{eq:SofGR-calLofGR}
reads~\cite{Weinberg1972,MisnerThorneWheeler2017,Wald1984}
\beq
\label{eq:Einstein-eq}
R_{\mu\nu}- \frac{1}{2}\,g_{\mu\nu}\,R= -8\pi G_N\, T_{\mu\nu}^{(M)}\,,
\eeq
where $R_{\mu\nu}$ is the Ricci curvature
tensor and $T_{\mu\nu}^{(M)}$ the energy-momentum tensor of the matter.
It is straightforward to evaluate \eqref{eq:Einstein-eq}
for the metric \eqref{eq:mod-RW-ds2}, as long as
the curvature tensors at $t=0$ are obtained by taking
the limit $t \to 0$; see Ref.~\cite{Guenther2017} for details.
If we also assume the energy-momentum tensor $T_{\mu\nu}^{(M)}$
of a homogeneous perfect fluid
[with energy density $\rho_{M}(t)$ and pressure $P_{M}(t)$
satisfying the standard energy conditions], then
modified Friedmann equations are obtained.
These modified Friedmann equations have been given
as Eqs.~(2.2) in Ref.~\cite{Klinkhamer2019-More}.

A heuristic understanding~\cite{Klinkhamer2019-More}
of these modified Friedmann equations
is that they can be rewritten
as the standard Friedmann equations with an
additional effective energy density $\rho_\text{defect}$
and an additional effective pressure $P_\text{defect}$,
both proportional to 
$-\widehat{b}^{\,2}\big/\big(\,\widehat{b}^{\,2}+t^2\big)$.
These effective quantities then violate the null energy condition
($\rho_\text{defect}+P_\text{defect} <0$) and allow for a bounce
of the cosmic scale factor $a(t)$
at $t=0$; see Ref.~\cite{NovelloBergliaffa2008} for a
general discussion on cosmic bounces and energy conditions.

But the modified Friedmann equations, as given by Eqs.~(2.2)
in Ref.~\cite{Klinkhamer2019-More}, can also be
written in another way:
\bsubeqs\label{eq:mod-Friedmann-equations-rewritten}
\beqa\label{eq:mod-Friedmann-equations-rewritten-1stFeq}
&&
\left( \frac{\dot{a}}{a}\right)^{2}
=\frac{8 \pi}{3}\,\widehat{G}_\text{eff}\,\rho_{M}\,,
\\[2mm]
\label{eq:mod-Friedmann-equations-rewritten-2ndFeq}
&&
\frac{\ddot{a}}{a}
+ \frac{1}{2}\,\left( \frac{\dot{a}}{a} \right)^2\,
=-4\pi \widehat{G}_\text{eff}\,P_{M}
+\frac{\widehat{b}^{\,2}}{\widehat{b}^{\,2}+t^{2}}\,\frac{1}{t}\,\frac{\dot{a}}{a}\,,
\\[2mm]
&&
\frac{d}{d a} \bigg[ a^{3}\,\rho_{M}(a)\bigg]+ 3\, a^{2}\,P_{M}(a)=0\,,
\\[3mm]
\label{eq:mod-Friedmann-equations-rewritten-Geff}
&&
\widehat{G}_\text{eff} =  \frac{t^2}{\widehat{b}^{\,2}+t^{2}}\,G_N\,,
\eeqa
\esubeqs
where the overdot stands for the derivative with respect to $t$.
The above equations have the same form as the standard
Friedmann equations~\cite{Weinberg1972},
except that Newton's gravitational coupling constant
$G_N$ is replaced by the time-dependent coupling 
$\widehat{G}_\text{eff}$ 
from \eqref{eq:mod-Friedmann-equations-rewritten-Geff}
and that there is an extra term on the right-hand side of
\eqref{eq:mod-Friedmann-equations-rewritten-2ndFeq}.
This extra term
makes the equations \eqref{eq:mod-Friedmann-equations-rewritten}
consistent for $\widehat{b}\ne 0$ and vanishes, formally, 
for $\widehat{b}=0$.

The basic idea, now, is to construct a Brans--Dicke-type
scalar-tensor theory~\cite{BransDicke1961},
which has a particular kink-type
solution of the scalar field that gives a behavior for
the effective gravitational coupling
similar to \eqref{eq:mod-Friedmann-equations-rewritten-Geff}.
In order to get this kink-type solution, the kinetic
term in the scalar field equation must have a ``wrong sign,''
which matches with the previous heuristic discussion about
an effective violation of the null energy condition.

\section{Scalar-tensor model}
\label{sec:Model}

The model to be presented in this section is not yet definitive.
Moreover, we do not have a local four-dimensional action
but only field equations. It is, of course, known that
certain ($d+1$)-dimensional theories exist without having 
a local ($d+1$)-dimensional action
(see Ref.~\cite{Witten1983} for a general discussion).
A model action will be given in  App.~\ref{sec:App-Brans-Dicke-Model},
but that action does not give the desired cosmology.
In fact, our main interest lies in the cosmological
equations, to be discussed in Sec.~\ref{subsec:Cosmological-equations}.
The incomplete model field equations of this section
are to be considered as a first step towards the appropriate
cosmological equations.

With a dimensionless real scalar field $\eta(x)$,
we define the model by its field equations,
\bsubeqs\label{eq:model-field-eqs}
\beqa
\label{eq:model-scalar-eq}
&&
b^{-2}\,\Box\, \eta = \lambda\,b^{-4}\,\big(1-\eta^{2}\big)\,\eta\,,
\\[2mm]
\label{eq:model-grav-eq}
&&
R_{\mu\nu}- \frac{1}{2}\,g_{\mu\nu}\,R
= -8\pi G\,\eta^2\, \left[ T_{\mu\nu}^{(\eta)}+T_{\mu\nu}^{(M)} \right]
+ X_{\mu\nu}\,,
\\[2mm]
\label{eq:model-Tmunu-eta}
&&
T_{\mu\nu}^{(\eta)}=
- b^{-2}\,\eta_{,\mu}\,\eta_{,\nu}
-g_{\mu\nu}\,
\left[
\frac{\lambda}{4}\, b^{-4}\,\big(1-\eta^{2}\big)^2
- \frac{1}{2}\,b^{-2}\,\eta_{,\lambda}\,\eta^{,\lambda}
\right]\,,
\\[2mm]
\label{eq:model-grav-eq-constants}
&& G >0\,, \quad \lambda>0 \,, \quad b >0\,,
\eeqa
\esubeqs
where $G$ is a gravitational coupling constant (mass dimension $-2$),
$\lambda$ a dimensionless quartic coupling constant (mass dimension $0$),
and $1/b$ a mass scale (mass dimension $1$).
In addition, we use the standard
notation~\cite{Weinberg1972,MisnerThorneWheeler2017}
of a comma for the derivative
and a semicolon for the covariant derivative
(for example, $\eta_{,\mu}\equiv \partial \eta/\partial x^\mu
\equiv \partial_\mu \eta$).
We keep the standard notation $\Box$ for the d'Alembertian,
$\Box\, \eta \equiv \eta_{,\mu}^{\;\;\; ;\,\mu}$
in the comma/semicolon notation for derivatives.

The gravitational field equation \eqref{eq:model-grav-eq}
contains the symmetric tensor $X_{\mu\nu}$ which is a functional of
the metric and the Brans--Dicke-type scalar field, and possibly also
the standard-matter fields $\psi$,%
\bsubeqs\label{eq:X}
\beqa\label{eq:X-functional}
&&X_{\mu\nu} =
X_{\mu\nu}\left[g_{\sigma\tau},\,  \partial_{\rho}g_{\sigma\tau},\,
\eta,\, \partial_{\rho}\eta,\,\psi,\,  \partial_{\rho}\psi,\,
\cdots\right]\,.
\eeqa
This tensor $X_{\mu\nu}$ has the purpose of
implementing energy-momentum conservation
in \eqref{eq:model-grav-eq},
\beqa
\label{eq:X-conservation}
&&\Big(
X_{\mu\nu}-8\pi G\,\eta^2\,
\left[ T_{\mu\nu}^{(\eta)}+T_{\mu\nu}^{(M)} \right]
\Big)^{;\nu}=0\,.
\eeqa
A further condition on $X_{\mu\nu}$ is that it vanishes
if the scalar field is in the vacuum configuration,%
\beqa
\label{eq:X-vanishing-in-eta-vacuum}
&&
X_{\mu\nu}\,\Big|_{\eta(x)=\pm 1}=0\,.
\eeqa
\esubeqs
For the moment, we will just assume that there exists
a proper expression for $X_{\mu\nu}$ which gives the
cosmological equations of Sec.~\ref{subsec:Cosmological-equations}
(a suggestion for a possible term $X_{\mu\nu}$ is
given in App.~\ref{sec:App-Possible-nonlocal-term}).
If such an $X_{\mu\nu}$ does not exist,
then the model needs to be extended.

Comparing the energy-momentum tensor \eqref{eq:model-Tmunu-eta} to
the one of a standard scalar field $\phi$,
we see that the two derivative terms in \eqref{eq:model-Tmunu-eta}
have ``wrong signs.''
The crucial point of this paper is that the
parameter $b$ appears already in the \emph{theory} \eqref{eq:model-field-eqs},
instead of only in the solution, as considered previously in \eqref{eq:mod-RW},
where the parameter was denoted $\widehat{b}$.

Obviously, we do not consider
the  model field equations \eqref{eq:model-field-eqs}
to describe a realistic theory, as we expect
instabilities and nonunitarity.
(Incidentally, the Pauli--Villars regulator fields of
quantum electrodynamics also have
wrong signs or wrong statistics~\cite{PauliVillars1949,ItzyksonZuber1980}.)
At best, the model field equations \eqref{eq:model-field-eqs}
would be embedded into a consistent UV completion
of Einstein's classical gravitation theory,
perhaps related to string theory.
For the moment, we only use the model
field equations \eqref{eq:model-field-eqs} to describe
the regularized big bang singularity,
without degenerate metrics but with
nonstandard matter fields (here, the scalar field $\eta$).

\section{Cosmology}
\label{sec:Cosmology}

\subsection{Ans\"{a}tze}
\label{subsec:Metric-Ansaetze}

We take the standard spatially flat Robertson--Walker (RW)
metric~\cite{Weinberg1972,MisnerThorneWheeler2017,Wald1984},
\bsubeqs\label{eq:RW}
\beqa\label{eq:RW-ds2}
\hspace*{-0mm}
ds^{2}
&=&
- d t^{2}+ a^{2}(t)\;\delta_{mn}\,dx^{m}\,dx^n\,,
\\[2mm]
\label{eq:RW-cosmic-scale-factor}
\hspace*{-0mm}
a(t) &\in& \mathbb{R}\,,
\\[2mm]
\label{eq:RW-range-cosmic-time-coordinate}
\hspace*{-0mm}
 t  &\in& (-\infty,\,\infty)\,,
\\[2mm]
\hspace*{-0mm}
x^{m} &\in& (-\infty,\,\infty)\,,
\eeqa
\esubeqs
where the cosmic time coordinate $t$ runs over the whole real line.
The metric from \eqref{eq:RW} is nondegenerate, as long as
$a(t) \ne 0$.

For the normal-matter content, we use again a homogeneous perfect fluid,
with energy density $\rho_{M}$ and pressure $P_{M}$
satisfying the standard energy conditions.
The ``wrong-sign'' Brans--Dicke-type scalar field $\eta$ is also taken
to be homogeneous. All these fields depend on cosmic time only,%
\bsubeqs\label{eq:homogeneous-matter-fields}
\beqa
\rho_{M} &=& \rho_{M}(t)\,,
\\[2mm]
P_{M}  &=& P_{M}(t)\,,
\\[2mm]
\eta
&=&
\eta(t)\,.
\eeqa
\esubeqs
We will now determine
the reduced  field equations from these \textit{Ans\"{a}tze}.

\subsection{Cosmological equations}
\label{subsec:Cosmological-equations}

With the metric \textit{Ansatz} \eqref{eq:RW}
and homogeneous matter fields \eqref{eq:homogeneous-matter-fields},
the scalar field equation \eqref{eq:model-scalar-eq} and
the 00 component of the
gravitational field equation \eqref{eq:model-grav-eq}
[under the assumption that $X_{00}$ effectively vanishes, see below]
reduce to two coupled ordinary differential equations (ODEs).
Added to these two ODEs are
the energy-conservation equation of the matter
and an equation of state $P_{M} = P_{M}\left( \rho_{M} \right)$
[here, we take a constant equation-of-state parameter
$w_{M}$]. All in all, the reduced equations are as follows:
\bsubeqs\label{eq:model-Friedmann-eqs}
\beqa
\hspace*{-8mm}
\label{eq:model-Friedmann-scalar-eq}
&& \ddot{\eta} + 3\,\frac{\dot{a}}{a} \, \dot{\eta} =
-\lambda\,b^{-2}\,\big(1-\eta^{2}\big)\,\eta\,,
\\[2mm]
\hspace*{-8mm}
\label{eq:model-Friedmann-adot-eq}
&&
\left( \frac{\dot{a}}{a} \right)^2
 =
\frac{8\pi}{3}\,G\,\eta^2\,
\left[ \rho_{M} - \frac{1}{2}\,b^{-2}\,\dot{\eta}^2
       + \frac{\lambda}{4}\, b^{-4}\,\big(1-\eta^{2}\big)^2 \right]\,,
\\[2mm]
\hspace*{-8mm}
\label{eq:model-Friedmann-rhoMdot-eq}
&&
\dot{\rho}_{M}+ 3\,\frac{\dot{a}}{a}\,\big(\rho_{M}+ P_{M}\big)=0\,,
\\[2mm]
\hspace*{-8mm}
&&
\label{eq:model-Friedmann-w-eq}
\frac{P_{M}(t)}{\rho_{M}(t) } = w_{M}  = \text{const}\,,
\eeqa
\esubeqs
where the last equation gives the equation of state
of the normal matter ($w_{M}=0$ for nonrelativistic matter
and $w_{M}=1/3$ for relativistic matter).
Observe the wrong sign of the
$\dot{\eta}^2$ term in the energy density
on the right-hand side of \eqref{eq:model-Friedmann-adot-eq}.

For the record, we mention that the 11 component of the
gravitational field equation \eqref{eq:model-grav-eq}
[with an appropriate expression for $X_{11}$, see below] gives
\beqa
\label{eq:model-Friedmann-11-grav-eq}
\frac{\ddot{a}}{a}
+ \frac{1}{2}\,\left( \frac{\dot{a}}{a} \right)^2
&=&
-4\pi G\,\eta^2\,
 \left[ P_{M} - \frac{1}{2}\,b^{-2}\,\dot{\eta}^2
        - \frac{\lambda}{4}\, b^{-4}\,\big(1-\eta^{2}\big)^2 \right]
+\frac{\dot{a}}{a}\,\frac{\dot{\eta}}{\eta}
\,,
\eeqa
but this equation is redundant.
With a given equation of state \eqref{eq:model-Friedmann-w-eq},
the ODE \eqref{eq:model-Friedmann-11-grav-eq}
can, in fact, be shown to follow from
the derivative of the ODE \eqref{eq:model-Friedmann-adot-eq},
by use of the ODEs \eqref{eq:model-Friedmann-scalar-eq},
\eqref{eq:model-Friedmann-adot-eq},
and \eqref{eq:model-Friedmann-rhoMdot-eq}.
This redundancy of the four ODEs is, in fact, a useful
check on their correctness and traces back to
the contracted Bianchi identities and energy-momentum
conservation (see Sec.~15.1 of Ref.~\cite{Weinberg1972}).

Note that the ODEs in \eqref{eq:model-Friedmann-eqs}
essentially have the standard
form [with wrong-sign kinetic terms of the scalar $\eta$
in \eqref{eq:model-Friedmann-scalar-eq}
and \eqref{eq:model-Friedmann-adot-eq} and
with an effective gravitational coupling $G\,\eta^2$ in the
first-order Friedmann equation \eqref{eq:model-Friedmann-adot-eq}],
but that \eqref{eq:model-Friedmann-11-grav-eq} has
an entirely new term $(\dot{a}/a)\,(\dot{\eta}/\eta)$
on the right-hand side.
This extra term traces back to the $X_{11}$ term in
the 11 component of the
original field equation \eqref{eq:model-grav-eq},
if the other reduced equations are used.
Equally, the term $X_{00}$ vanishes effectively.
In short, we suppose that, with the \textit{Ans\"{a}tze}
of Sec.~\ref{subsec:Metric-Ansaetze} and the
reduced field equations of this subsection, we have
\bsubeqs\label{eq:Xmunu-reduced}
\beqa
\label{eq:Xmunu-reduced-00}
X_{00}\,\Big|_\text{reduced} &\stackrel{\circ}{=}& 0\,,
\\[2mm]
\label{eq:Xmunu-reduced-mm}
X_{m  m}\,\Big|_\text{reduced}
&\stackrel{\circ}{=}&
2\,a^2\; \frac{\dot{a}}{a}\,\frac{\dot{\eta}}{\eta}\,,
\\[2mm]
\label{eq:Xmunu-reduced-off-diag}
X_{\mu\nu}\,\Big|_\text{reduced} &\stackrel{\circ}{=}& 0\,,
\quad \text{for}\;\; \mu\ne\nu\,,
\eeqa
\esubeqs
where the spatial index ``$m$'' on the left-hand side of \eqref{eq:Xmunu-reduced-mm} is not summed over and
where the symbol ``$\stackrel{\circ}{=}$'' indicates that the equality
holds only ``on-shell.''

The cosmological equations \eqref{eq:model-Friedmann-eqs}
and \eqref{eq:model-Friedmann-11-grav-eq}  are the main general result of
this paper.

\section{Cosmological solutions}
\label{sec:Cosmological-solutions}

\subsection{Constant EOS parameter}
\label{subsec:Constant-EOS-parameter}

In the following, we use a constant equation-of-state (EOS)
parameter \eqref{eq:model-Friedmann-w-eq}. Then,
the solution of \eqref{eq:model-Friedmann-rhoMdot-eq}
is explicitly given by
\bsubeqs\label{eq:rhoMsol-r0-nonnegative}
\beqa
\label{eq:rhoMsol}
\rho_{M}(a) &=& \rho_{M0}\;a^{-3\,\left(1+w_{M}\right)}\,,
\\[2mm]
\label{eq:r0-nonnegative}
\rho_{M0} &\geq& 0\,,
\eeqa
\esubeqs
where $a=a(t)$ is assumed to be positive.
This implies that there are essentially two dimensionless
functions to be determined, 
namely $\eta(t)$ and $a(t)$.
Solving the ODEs \eqref{eq:model-Friedmann-eqs} with
different boundary conditions for these two functions
results, of course, in different solutions.

\subsection{Dimensionless ODEs}
\label{subsec:Dimensionless-ODEs}

At this moment, it turns out to be useful to introduce the following
dimensionless quantities (recall $c=\hbar=1$):%
\bsubeqs\label{eq:dimensionless-quantities}
\beqa
\label{eq:dimensionless-quantities-g}
g &\equiv& G/b^2\,,
\\[2mm]
\label{eq:dimensionless-quantities-tau}
\tau &\equiv& t/b\,,
\\[2mm]
\label{eq:dimensionless-quantities-rM}
r_{M}(\tau) &=&  \rho_{M}(t)\,b^4\,,
\eeqa
\esubeqs
together with, by the usual abuse of notation,
$a(\tau)=a(t)$ and $\eta(\tau)=\eta(t)$.
From now on, there is no danger of misunderstanding
the meaning of $g$, as the determinant of the metric no longer appears.

Introducing dimensionless quantities, we obtain
from \eqref{eq:model-Friedmann-eqs} and \eqref{eq:rhoMsol-r0-nonnegative}
the following dimensionless cosmological equations:
\bsubeqs\label{eq:dimensionless-model-Friedmann-eqs}
\beqa
\hspace*{-8mm}
\label{eq:dimensionless-model-Friedmann-scalar-eq}
&&
\ddot{\eta} + 3\,\frac{\dot{a}}{a} \, \dot{\eta} =
-\lambda\,\big(1-\eta^{2}\big)\,\eta\,,
\\[3mm]
\hspace*{-8mm}
\label{eq:dimensionless-model-Friedmann-adot-eq}
&&
\left( \frac{\dot{a}}{a} \right)^2
 =
\frac{8\pi}{3}\,g\,\eta^2\,
\left[ r_{M} - \frac{1}{2}\,\dot{\eta}^2
+ \frac{\lambda}{4}\, \big(1-\eta^{2}\big)^2 \right]\,,
\\[3mm]
\hspace*{-8mm}
\label{eq:dimensionless-model-Friedmann-rhoM-sol}
&&
r_{M} = r_{M0}\;a^{-3\,\left(1+w_{M}\right)}\,,
\eeqa
\esubeqs
where the overdot now stands for the derivative with respect to $\tau$,
$w_{M}$ is the constant EOS parameter, and $r_{M0}\geq 0$
is a constant coming from the boundary conditions (see below).
The dimensionless 11-component gravitational equation
from \eqref{eq:model-Friedmann-11-grav-eq} reads
\beqa
\label{eq:dimensionless-model-Friedmann-11-grav-eq}
\frac{\ddot{a}}{a}
+ \frac{1}{2}\,\left( \frac{\dot{a}}{a} \right)^2
&=&
-4\pi g\,\eta^2\,
 \left[ w_{M}\,r_{M} - \frac{1}{2}\,\dot{\eta}^2
        - \frac{\lambda}{4}\, \big(1-\eta^{2}\big)^2 \right]
        +\frac{\dot{a}}{a}\,\frac{\dot{\eta}}{\eta}\,,
\eeqa
with $r_{M}$ given by \eqref{eq:dimensionless-model-Friedmann-rhoM-sol}.

\subsection{Scalar vacuum solution without matter}
\label{subsec:Scalar-vacuum-solution-without-matter}

We take the following boundary conditions:
\bsubeqs\label{eq:bcs-scalar-vacuum-without-matter}
\beqa
\label{eq:bcs-scalar-vacuum-without-matter-eta-at-t-is-plus-infty}
\lim_{\tau \to \infty}\eta(\tau)&=& 1\,,
\\[2mm]
\label{eq:bcs-scalar-vacuum-without-matter-eta-at-t-is-zero}
\eta(0)&=& 1\,,
\\[2mm]
\label{eq:bcs-scalar-vacuum-without-matter-a-at-t-is-zero}
a(0)&=& 1 \,,
\\[2mm]
\label{eq:bcs-scalar-vacuum-without-matter-rhoM-at-t-is-zero}
r_{M}(0)&=& 0\,.
\eeqa
\esubeqs
The solution of the ODEs \eqref{eq:dimensionless-model-Friedmann-eqs} and
\eqref{eq:dimensionless-model-Friedmann-11-grav-eq} is then as follows:
\bsubeqs\label{eq:sol-scalar-vacuum-without-matter}
\beqa
\label{eq:sol-scalar-vacuum-without-matter-eta}
\eta(\tau)&=& 1\,,
\\[2mm]
\label{eq:sol-scalar-vacuum-without-matter-a}
a(\tau)&=& 1 \,,
\\[2mm]
\label{eq:sol-scalar-vacuum-without-matter-rhoM}
r_{M}(\tau)&=& 0\,,
\eeqa
\esubeqs
which corresponds to an empty static universe with
the Minkowski metric and the $\eta$-vacuum.

\subsection{Scalar vacuum solution with relativistic matter}
\label{subsec:Scalar-vacuum-solution-with-relativistic-matter}

We, now, consider relativistic matter with equation-of-state parameter
\beq
\label{eq:wM-relativistic}
w_{M}=1/3\,.
\eeq
Other values $w_{M} \in [0 ,\,1]$ are certainly possible,
but, for definiteness, we focus on $w_{M}=1/3$ in the following.
In addition, we take the boundary conditions
\bsubeqs\label{eq:bcs-scalar-vacuum-with-rel-matter}
\beqa
\label{eq:bcs-scalar-vacuum-with-rel-matter-eta-at-t-is-plus-infty}
\lim_{\tau \to \infty}\eta(\tau)&=& 1\,,
\\[2mm]
\label{eq:bcs-scalar-vacuum-with-rel-matter-eta-at-t-is-zero}
\eta(\tau_{0})&=& 1\,,
\\[2mm]
\label{eq:bcs-scalar-vacuum-with-rel-matter-a-at-t-is-zero}
a(\tau_{0})&=& 1 \,,
\\[2mm]
\label{eq:bcs-scalar-vacuum-with-rel-matter-rhoM-at-t-is-zero}
r_{M}(\tau_{0})&=& r_{M0} > 0\,,
\eeqa
\esubeqs
for a fixed finite time $\tau_{0}>0$. Alternative boundary conditions
for the Brans--Dicke-type
scalar field would be $\eta(\tau_{0})=1$ and $\dot{\eta}(\tau_{0})=0$,
but the boundary conditions 
\eqref{eq:bcs-scalar-vacuum-with-rel-matter-eta-at-t-is-plus-infty}
and \eqref{eq:bcs-scalar-vacuum-with-rel-matter-eta-at-t-is-zero}
are more suitable for the comparison with the boundary conditions
in Sec.~\ref{subsec:Scalar-kink-solution-with-relativistic-matter}.

\begin{figure}[t!]
\vspace*{-0mm}
\begin{center}
\includegraphics[width=0.625\textwidth]{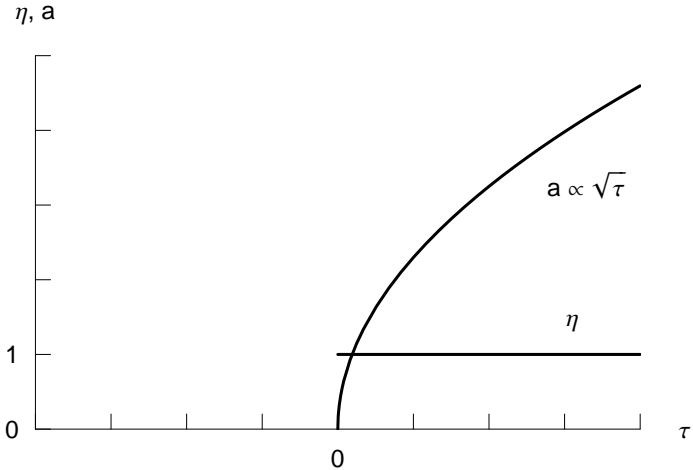}
\end{center}
\vspace*{-5mm}
\caption{Qualitative behavior of the FLRW solution of the
cosmological equations \eqref{eq:dimensionless-model-Friedmann-eqs}
and \eqref{eq:dimensionless-model-Friedmann-11-grav-eq}:
the horizontal curve shows the constant Brans--Dicke-type
scalar field $\eta(\tau)=1$
and the rising curve shows the cosmic scale factor $a(\tau)$ for $\tau>0$.
Relativistic matter is assumed to be present
and its energy density is given by $r_{M}(\tau) =r_{M0}/a(\tau)^4$
for $r_{M0}>0$.}
\label{fig:fig01}
\vspace*{0mm}
\end{figure}

The solution of the ODEs \eqref{eq:dimensionless-model-Friedmann-eqs} and
\eqref{eq:dimensionless-model-Friedmann-11-grav-eq}
with  boundary conditions \eqref {eq:bcs-scalar-vacuum-with-rel-matter}
is as follows:
\bsubeqs\label{eq:sol-scalar-vacuum-with-rel-matter}
\beqa
\label{eq:sol-scalar-vacuum-with-rel-matter-eta}
\eta(\tau)&=& 1\,,
\\[2mm]
\label{eq:bcs-scalar-vacuum-with-rel-matter-a}
a(\tau)&=& \sqrt{\frac{\tau}{\tau_{0}}} \,,
\\[2mm]
\label{eq:bcs-scalar-vacuum-with-rel-matter-rhoM}
r_{M}(\tau)&=& r_{M0}\;\frac{\tau_{0}^2}{\tau^2}\,,
\eeqa
\esubeqs
which corresponds to the standard
Friedmann--Lema\^{i}tre--Robertson--Walker (FLRW)
universe~\cite{Friedmann1922-1924,Weinberg1972,%
MisnerThorneWheeler2017,Wald1984}.
The solution \eqref{eq:sol-scalar-vacuum-with-rel-matter}
holds only for $\tau>0$ and there is a big bang singularity
at $\tau=0$ with diverging curvature and energy density.
Figure~\ref{fig:fig01} gives a sketch, in order to prepare
for the comparison with the solution of
Sec.~\ref{subsec:Scalar-kink-solution-with-relativistic-matter}.
Here, we observe that the $a(\tau)$ curve in Fig.~\ref{fig:fig01}
is concave, with $\ddot{a}(\tau)/a(\tau)<0$
from the ODE \eqref{eq:dimensionless-model-Friedmann-11-grav-eq}
for $\eta(\tau)=1$.

\subsection{Scalar kink solution with relativistic matter}
\label{subsec:Scalar-kink-solution-with-relativistic-matter}

\subsubsection{Boundary conditions}
\label{subsubsec:kink-solution-with-rel-matter-Boundary-conditions}

In order to ``tame'' the big bang singularity of
Sec.~\ref{subsec:Scalar-vacuum-solution-with-relativistic-matter},
we suggest to use a nontrivial Brans--Dicke-type scalar field $\eta(t)$.
Specifically, we consider again
relativistic matter with the equation-of-state parameter
\beq
\label{eq:wM-relativistic-again}
w_{M}=1/3\,,
\eeq
but now take the following boundary conditions:
\bsubeqs\label{eq:bcs-scalar-kink-with-rel-matter}
\beqa
\label{eq:bcs-scalar-kink-with-rel-matter-eta-at-t-is-plus-infty}
\lim_{\tau \to \infty}\eta(\tau)&=& 1\,,
\\[2mm]
\label{eq:bcs-scalar-kink-with-rel-matter-eta-at-t-is-zero}
\eta(0)&=& 0\,,
\\[2mm]
\label{eq:bcs-scalar-kink-with-rel-matter-a-at-t-is-zero}
a(0)&=& 1 \,,
\\[2mm]
\label{eq:bcs-scalar-kink-with-rel-matter-rhoM-at-t-is-zero}
r_{M}(0)&=& r_{M0} > 0\,.
\eeqa
\esubeqs
The boundary condition \eqref{eq:bcs-scalar-kink-with-rel-matter-eta-at-t-is-zero}
rules out the trivial solution $\eta(t)= 1$
and fixes the core of the temporal-kink solution to be at $\tau=0$
(see Sec.~\ref{subsubsec:Exact-flat-spacetime-solution} for details).
The boundary condition $\lim_{\tau \to -\infty}\eta(\tau)= -1$
would also rule out the trivial solution but would leave
the core location of the temporal-kink solution free.

\subsubsection{Exact flat-spacetime solution}
\label{subsubsec:Exact-flat-spacetime-solution}

As a start, we give an exact solution for the case of no
direct gravitational interaction ($G=0$).
Using dimensionless variables, this temporal-kink solution
of the ODEs \eqref{eq:dimensionless-model-Friedmann-scalar-eq},
\eqref{eq:dimensionless-model-Friedmann-adot-eq},
and \eqref{eq:dimensionless-model-Friedmann-11-grav-eq}
is given by
\bsubeqs\label{eq:exact-sol-g-zero}
\beqa
\label{eq:exact-sol-g-zero-eta}
\eta_\text{exact-sol}^{(g=0)}(\tau)
&=& \tanh \left(\sqrt{\lambda/2}\;\tau\right)\,,
\\[2mm]
\label{eq:exact-sol-g-zero-a}
a_\text{exact-sol}^{(g=0)}(\tau)&=& 1\,.
\eeqa
\esubeqs
The matter content is irrelevant for this static
cosmology, as its gravitational interaction vanishes.

The time-dependent homogeneous scalar
configuration \eqref{eq:exact-sol-g-zero-eta}
has, by construction, the same mathematical structure as the 
static finite-energy kink solution of the quartic scalar theory
(see Chap. 6, Sec. 2.1 in Ref.~\cite{Coleman1985}
and Chap. 5, Sec. 5.2 in Ref.~\cite{MantonSutcliffe2004}).
As the standard kink solution [or, rather, the domain-wall solution
in the (3+1)-dimensional context] involves a single spatial coordinate,
say $x^1$, it is clear that a similar solution involving the
time coordinate $x^0$ requires a theory with a ``wrong-sign''
kinetic term. Such a ``wrong-sign'' theory has, of course,
no longer the standard static kink (or  domain-wall) solution.

The solution for $g>0$ that
we will discuss in the rest of this section can be viewed as a
\emph{deformation} of the solution \eqref{eq:exact-sol-g-zero}.
By a continuity argument, we expect such a solution
to exist over an interval $[-\tau_\text{max},\,\tau_\text{max}]$
for small enough values of $g>0$, $r_{M0} \geq 0$, and $\tau_\text{max}>0$.

\subsubsection{Series solution}
\label{subsubsec:Series-solution}

From the ODEs \eqref{eq:dimensionless-model-Friedmann-eqs}
for general values of $w_{M}$,
we obtain the following series solution near $\tau=0$:
\bsubeqs\label{eq:pert-sol}
\beqa
\label{eq:pert-sol-1-eta}
\eta_\text{pert-sol}(\tau)&=& e_1\;\tau+ e_3\;\tau^3+ e_5\;\tau^5 + \cdots\,,
\\[2mm]
\label{eq:pert-sol-a}
a_\text{pert-sol}(\tau)
&=&
1 + a_2\;\tau^2 + a_4\;\tau^4 + \cdots\,,
\\[2mm]
r_{M\text{pert-sol}}(\tau)&=&
r_{M0}\;[a_\text{pert-sol}(\tau)]^{-3\,\left(1+w_{M}\right)}\,,
\eeqa
\esubeqs
with coefficients
\bsubeqs\label{eq:pert-sol-coeff}
\beqa
e_1 &=& \sqrt{\lambda/2}\,,
\\[2mm]
e_3 &=&
-\frac{\lambda}{6}\,
\left(
 \sqrt{\lambda/2} +  \sqrt{6\,\pi\,g\,r_{M0}}
\right)\,,
\\[2mm]
e_5 &=&
\frac{\lambda^{3/2}}{120\, {\sqrt{2}}}\,
\left(
\lambda\, \left[ 4 - 9\, g\, \pi  \right]
+ 10\,\sqrt{3\, \pi\,\lambda\,g\,r_{M0}}
+ 6\,  \pi \,\left[ 11 + 3\, w_{M} \right]\,g\, r_{M0}
  \right)\,,
\\[2mm]
a_2 &=&{\sqrt{\pi/3}}\, \sqrt{\lambda\,g\,r_{M0}}\,,
\\[2mm]
a_4 &=&
-\frac{\lambda}{72}\,
\left(
2\,{\sqrt{3\,\pi\,\lambda\,g\,r_{M0}}}
- 9\, g\,  \pi \,
\left[ \lambda - 2\, r_{M0}\, \left( 1 + w_{M} \right)  \right] \right)\,,
\eeqa
\esubeqs
where we have chosen the positive root for $a_2$.
For $g\,r_{M0} \lesssim \lambda$, the above series
are essentially perturbation expansions in $\lambda$.

Observe that, even in the absence of normal matter,
the spacetime is curved:
\bsubeqs\label{eq:pert-sol-coeff-no-matter}
\beqa
a_2\,\Big|_{r_{M0}=0} &=& 0\,,
\\[2mm]
a_4\,\Big|_{r_{M0}=0} &=& (\pi/8)\, g\, \lambda^2\,,
\eeqa
\esubeqs
where the last coefficient does not vanish, as long as both
$g$ and $\lambda$ are nonzero. Incidentally, the coefficients
$a_2$ and $a_4$  do vanish if $g=0$, consistent with the
exact solution \eqref{eq:exact-sol-g-zero}.

The kink-bounce series solution \eqref{eq:pert-sol}
and \eqref{eq:pert-sol-coeff} is the main analytic result of
this paper.

\subsubsection{Approximate solution}
\label{subsubsec:Approximate-solution}

In addition to the series solution, we have the following approximate
solution for $w_{M}=1/3$ (similar results have been obtained for other
values of $w_{M}$):
\bsubeqs\label{eq:approx-sol}
\beqa\label{eq:approx-sol-eta}
\eta_\text{approx-sol}(\tau)&=& \tanh \left(\sqrt{\lambda/2}\;\tau\right)\,,
\\[2mm]
\label{eq:approx-sol-a}
a_\text{approx-sol}(\tau) &=&
\sqrt{
1 + 8\,\sqrt{\pi/3}\,\sqrt{g\,r_{M0}/\lambda}\;
        \ln \left[\cosh \left(\sqrt{\lambda/2}\,\tau\right)\right]
}  \,,
\\[2mm]
\label{eq:approx-sol-rM}
r_{M\text{-approx-sol}}(\tau) &=& r_{M0}\;[a_\text{approx-sol}(\tau)]^{-4}\,,
\eeqa
\esubeqs
where we have chosen the expanding branch for $\tau > 0$.
The configurations \eqref{eq:approx-sol-a}
and \eqref{eq:approx-sol-rM}
provide an exact solution of the first-order Friedmann equation
\eqref{eq:dimensionless-model-Friedmann-adot-eq}
for the given scalar function \eqref{eq:approx-sol-eta}.
But \eqref{eq:approx-sol-eta}
is only an approximate solution of the second-order scalar
equation \eqref{eq:dimensionless-model-Friedmann-scalar-eq},
due to the uncancelled  ``friction'' term $3\,(\dot{a}/a) \,\dot{\eta}$.
Indeed, the Taylor coefficients from \eqref{eq:approx-sol-eta}
do not involve $r_{M0}$, whereas the coefficients $e_3$ and $e_5$
of the genuine solution do have a nontrivial dependence on $r_{M0}$,
according to the results  \eqref{eq:pert-sol-coeff}.
Equally, the cosmic scale factor configuration \eqref{eq:approx-sol-a}
provides only an approximative solution, as it has
$a_\text{approx-sol}(\tau)=1$ for $r_{M0}=0$, whereas
the genuine solution for $r_{M0}=0$ still has a nontrivial time dependence
of the cosmic scale factor, according to \eqref{eq:pert-sol-coeff-no-matter}.

\begin{figure}[t!]  
\vspace*{-0mm}
\begin{center}
\includegraphics[width=0.625\textwidth]{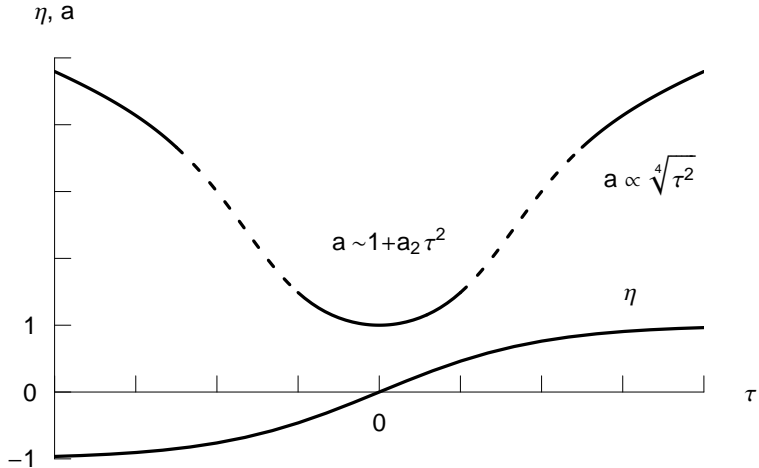}
\end{center}
\vspace*{-0mm}
\caption{Qualitative behavior of the kink-bounce solution
of the cosmological equations \eqref{eq:dimensionless-model-Friedmann-eqs}
and \eqref{eq:dimensionless-model-Friedmann-11-grav-eq}:
the bottom curve shows the kink configuration of the Brans--Dicke-type
scalar field $\eta(\tau)$
and the top curve shows the cosmic scale factor $a(\tau)$ with
a bounce at $\tau=0$. The zero of the $\eta(\tau)$ function
sets the moment of the bounce [stationary point of the curve $a(\tau)$].
Relativistic matter is assumed to be present
and its energy density is given by $r_{M}(\tau) =r_{M0}/a(\tau)^4$
for $r_{M0}>0$.}
\label{fig:fig02}
\vspace*{0mm}
\end{figure}

For the record, the corresponding dimensionless Hubble variable
($h \equiv \dot{a}/a$) reads:%
\beq
h_\text{approx-sol}(\tau) =
\frac{ 2\,\sqrt{2\,\pi/3}\,\sqrt{g\,r_{M0}}\;
    \tanh\left({\sqrt{\lambda/2}}\;\tau\right)}
    {1 + 8\,\sqrt{\pi/3}\,\sqrt{g\,r_{M0}/\lambda}\;
        \ln \left[\cosh \left(\sqrt{\lambda/2}\;\tau\right)\right]}\,.
\eeq
The cosmic scale factor and the Hubble variable
near $\tau=0$ are, now, given by%
\bsubeqs\label{eq:a-h-approxsol-series}
\beqa
a_\text{approx-sol}(\tau) &=&
1 + \sqrt{\pi/3}\,
   {\sqrt{g\,\lambda\,r_{M0}}}\,{\tau}^2 +\text{O}(\tau^4)\,,
\\[2mm]
h_\text{approx-sol}(\tau) &=& 2\,\sqrt{\pi/3}\,
   {\sqrt{g\,\lambda\,r_{M0}}}\,{\tau} +\text{O}(\tau^3)\,.
\eeqa
\esubeqs
As $|\tau| \to \infty$, these functions are given by
\bsubeqs\label{eq:a-h-approxsol-asymp}
\beqa
a_\text{approx-sol}(\tau) &\propto&  \left(\tau^2\right)^{1/4}\,,
\\[2mm]
h_\text{approx-sol}(\tau) &\sim& \frac{1}{2}\,\tau^{-1}\,.
\eeqa
\esubeqs
For the numerics later on, we will
use  the approximate solution \eqref{eq:approx-sol}
at $\tau \gg 1$, where the approximate solution is close to the
exact FLRW solution \eqref{eq:sol-scalar-vacuum-with-rel-matter}.
Remark also that, with $\eta^2$ from
\eqref{eq:exact-sol-g-zero-eta} approaching unity exponentially fast
as $|\tau| \to \infty$, it is possible to identify
$G$ with Newton's  gravitational coupling constant $G_N$,
first measured by Cavendish.

At this moment, we observe two different time scales
in the perturbative solution \eqref{eq:pert-sol},
one for the scalar field $\eta$
and another for the cosmic scale factor $a$:
\bsubeqs\label{eq:pert-sol-time-scales}
\beqa
\tau_\text{scale}^{(\eta)}  &\equiv&  \left( \lambda\right)^{-1/2}\,,
\\[2mm]
\tau_\text{scale}^{(a)}     &\equiv& \left( \lambda\,g\,r_{M0} \right)^{-1/4} \,,
\eeqa
\esubeqs
where we have assumed that $r_{M0} \lesssim \lambda$ and $g \lesssim 1$.
We expect that the perturbative  solution
\eqref{eq:pert-sol} changes
to the approximate solution \eqref{eq:approx-sol}
around $|\tau| \sim \sqrt{2/\lambda}$ with $\eta \sim 1/2$
(see Fig.~\ref{fig:fig02} for a sketch and compare with
Fig.~\ref{fig:fig01}).
Remark that the $a(\tau)$ curve in Fig.~\ref{fig:fig01} is convex at $\tau=0$,
with $\ddot{a}(0)/a(0)>0$
from the term $(\dot{a}/a)\,(\dot{\eta}/\eta)$ on the right-hand side
of \eqref{eq:dimensionless-model-Friedmann-11-grav-eq}.

\begin{figure}[t]  
\vspace*{-0mm}
\begin{center}
\includegraphics[width=1\textwidth]{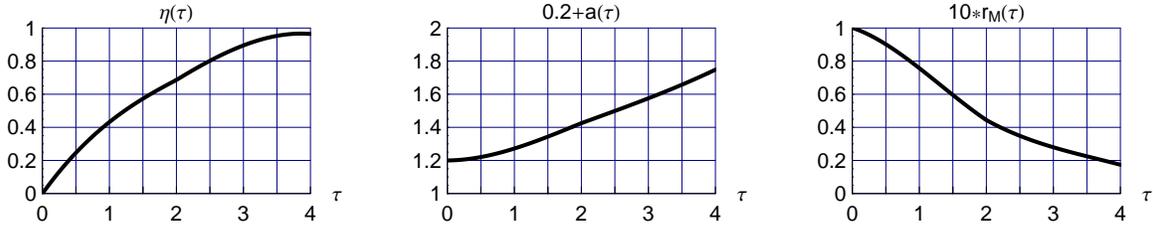}
\end{center}
\vspace*{-0mm}
\caption{Numerical solution of the cosmological ODEs
\eqref{eq:dimensionless-model-Friedmann-eqs}
and \eqref{eq:dimensionless-model-Friedmann-11-grav-eq}
over the time interval $[\tau_\text{min},\,\tau_\text{max}]$,
with  $\tau_\text{min}=1/100$ and $\tau_\text{max}=4$.
The boundary conditions at $\tau=\tau_\text{min}$ follow from the
perturbative solution \eqref{eq:pert-sol} and \eqref{eq:pert-sol-coeff},
while the boundary conditions
at $\tau=\tau_\text{max}$ follow from the approximate solution
\eqref{eq:approx-sol}.
The model parameters are $g=1/10$, $\lambda=1/2$, and $w_{M}=1/3$.
The matter density at the moment of the bounce is $r_{M0}=1/10$.
}
\label{fig:fig03}
\vspace*{0mm}
\end{figure}

\subsubsection{Numerical solution}
\label{subsubsec:Scalar-kink--rel-matter-Numerical-solution}

We have obtained the numerical solution of the second-order
ODEs \eqref{eq:dimensionless-model-Friedmann-scalar-eq}
and \eqref{eq:dimensionless-model-Friedmann-11-grav-eq}
over the cosmic time interval $[\tau_\text{min},\,\tau_\text{max}]$,
for  $\tau_\text{max} > \tau_\text{min} >0$.
With boundary conditions at $\tau=\tau_\text{min}$ from the
perturbative solution
\eqref{eq:pert-sol} and \eqref{eq:pert-sol-coeff},
we numerically integrate forward in time.
With boundary conditions
at $\tau=\tau_\text{max}$ from the approximate solution
\eqref{eq:approx-sol},  we numerically integrate backward in time.
These two numerical solutions are then matched at an appropriate
intermediate time $\tau_\text{match} \in [\tau_\text{min},\,\tau_\text{max}]$.
The boundary conditions at $\tau=\tau_\text{min}$ and $\tau=\tau_\text{max}$
satisfy the first-order
ODE \eqref{eq:dimensionless-model-Friedmann-adot-eq}
and the residue of this ODE can be used to monitor the accuracy
of the numerical solution at intermediate time values.

Figure~\ref{fig:fig03} shows the numerical solution
for a particular choice of model parameters $g$, $\lambda$, and $w_{M}$,
and for a particular value of the matter density $r_{M0}$ 
at the moment of the bounce.

\section{Discussion}\label{sec:Discussion}

In the present article, we have presented a
somewhat baroque model with cosmological equations
that may have a nonsingular bounce solution.
The bounce solution appears if the boundary
conditions allow for a kink-type solution of
the ``wrong-sign''Brans--Dicke-type scalar field.  If the boundary
conditions allow only for a vacuum-type solution of
the scalar field, then the standard Friedmann solution
is recovered, which has a big bang curvature
singularity. It is rather interesting that a single
model, regardless of how baroque, allows for both types
of behavior (sketched in Figs.~\ref{fig:fig01} and \ref{fig:fig02}).
The time scale $b/c$ that describes
the structure of the bounce is set by the model
field equations and not by the \textit{Ansatz} for the
solution, as is the case for the
degenerate-metric bounce~\cite{Klinkhamer2019},
where the \textit{Ansatz} parameter was denoted $\widehat{b}$.

Even with a different origin of the parameters $b$ and $\widehat{b}$,
the scalar-model cosmological equations \eqref{eq:model-Friedmann-eqs}
and \eqref{eq:model-Friedmann-11-grav-eq}
are quite similar to the degenerate-metric cosmological equations
\eqref{eq:mod-Friedmann-equations-rewritten}.
The interpretation is that these scalar-model cosmological equations
provide a dynamic realization of the effective gravitational
coupling that was obtained from a degenerate-metric
\textit{Ansatz} in general relativity.
Specifically, the scalar-model bounce has $G_\text{eff}=\eta^2\,G$,
which, for the kink-type configuration \eqref{eq:exact-sol-g-zero-eta}, 
matches the expression \eqref{eq:mod-Friedmann-equations-rewritten-Geff} 
of the degenerate-metric bounce.

We need to mention one important open problem,
namely the study of perturbations and stability of the
kink-bounce solution of Sec.~\ref{subsec:Scalar-kink-solution-with-relativistic-matter}.
For the degenerate-metric bounce, these issues were
addressed in Ref.~\cite{KlinkhamerWang2020-PRD}.
Some preliminary results on the linear perturbations of the kink-bounce
solution are presented in App.~\ref{sec:App-Linear-perturbations}.

In closing, we have two general remarks. First, we note that
the ``future'' boundary condition
\eqref{eq:bcs-scalar-kink-with-rel-matter-eta-at-t-is-plus-infty}
of the kink-bounce solution
is somewhat surprising in view of the
other ``initial'' boundary conditions
\eqref{eq:bcs-scalar-kink-with-rel-matter-eta-at-t-is-zero},
\eqref{eq:bcs-scalar-kink-with-rel-matter-a-at-t-is-zero},
and \eqref{eq:bcs-scalar-kink-with-rel-matter-rhoM-at-t-is-zero}.
But two-times boundary conditions may be less amazing than
they appear at first sight; see Ref.~\cite{Shulman1997} for
further discussion.

Second, the present article has considered a modified gravity
theory with a nonstandard matter field and a mass parameter $1/b$.
Perhaps it is also possible to construct a geometric version of
the modified gravity theory with a length parameter $\widetilde{b}$.
Such a new geometric theory, if it exists,
may or may not be approximated by general relativity with degenerate metrics.

\begin{acknowledgments}
It is a pleasure to thank E. Battista and Z.L. Wang
for comments on the manuscript.
\end{acknowledgments}

\begin{appendix}

\section{Brans--Dicke model}
\label{sec:App-Brans-Dicke-Model}

\subsection{Action and field equations}
\label{subsec:Action-and-field-equations}

With a dimensionless real scalar field $\eta(x)$,
the Brans--Dicke action~\cite{BransDicke1961} is taken as follows:%
\bsubeqs\label{eq:app-Smodel-calLmodel}
\beqa\label{eq:app-Smodel}
\widetilde{S} &=& \int d^4 x \,\sqrt{-g}  \,\widetilde{\mathcal{L}}\,,
\\[2mm]
\label{eq:app-calLmodel}
\widetilde{\mathcal{L}}&=&
-\frac{1}{16 \pi G}\,\frac{1}{\eta^2}\,R
+ \frac{1}{2}\, b^{-2}\, \partial_\mu  \eta \, \partial^\mu \eta
-\frac{\lambda}{4}\, b^{-4}\,\big(1-\eta^{2}\big)^2
+ \mathcal{L}_{M}\,,
\\[2mm]
G &>&0\,, \quad \lambda>0 \,, \quad b >0\,,
\eeqa
\esubeqs
where $G$ is a gravitational coupling constant and $1/b$ a mass scale.
With a negative metric component $g_{00}$, we see that the 
sign of the term $(\partial_t \eta)^2$
from the scalar kinetic term in \eqref{eq:app-calLmodel}
equals the sign of the scalar potential term, so that
we have a ``wrong-sign'' kinetic term for $\eta$
in our model action (recall the classical-mechanics expression
$L=T-V$, with $L$ the Lagrangian, $T$ the kinetic energy,
and $V$ the potential energy).

The model field equations from \eqref{eq:app-Smodel-calLmodel} are:
\bsubeqs\label{eq:app-model-field-eqs}
\beqa
\label{eq:app-model-scalar-eq}
&&
b^{-2}\,\Box\, \eta = \lambda\,b^{-4}\,\big(1-\eta^{2}\big)\,\eta +
\frac{1}{8 \pi G}\,\frac{1}{\eta^3}\,R\,,
\\[2mm]
\label{eq:app-model-grav-eq}
&&
R_{\mu\nu}- \frac{1}{2}\,g_{\mu\nu}\,R
= -8\pi G\,\eta^2\, \left[ T_{\mu\nu}^{(\eta)}+T_{\mu\nu}^{(M)} \right]
-\eta^2\,\left[
\left(\eta^{-2}\right)_{,\mu ;\nu}
- g_{\mu\nu}\,\Box\,\left(\eta^{-2}\right)\right]
\,,
\\[2mm]
&&
T_{\mu\nu}^{(\eta)}=
- b^{-2}\,\eta_{,\mu}\,\eta_{,\nu}
-g_{\mu\nu}\,
\left[
\frac{\lambda}{4}\, b^{-4}\,\big(1-\eta^{2}\big)^2
- \frac{1}{2}\,b^{-2}\,\eta_{,\lambda}\,\eta^{,\lambda}
\right]
\,,
\eeqa
\esubeqs
where we use the standard
notation~\cite{Weinberg1972,MisnerThorneWheeler2017}
of a comma for the derivative
and a semicolon for the covariant derivative.

Just as for the model of Sec.~\ref{sec:Model}, we do not consider
the  model action \eqref{eq:app-Smodel-calLmodel}
to describe a realistic theory, as we expect
instabilities and nonunitarity.

\subsection{Cosmological equations}
\label{subsec:App-Cosmological-equations}

In the cosmological context,
we now make the same \textit{Ans\"{a}tze} as
in Sec.~\ref{subsec:Metric-Ansaetze}.
With the metric \textit{Ansatz} \eqref{eq:RW}
and homogeneous matter fields \eqref{eq:homogeneous-matter-fields},
the scalar field equation \eqref{eq:app-model-scalar-eq}
and the 00 component of the
gravitational field equation \eqref{eq:app-model-grav-eq}
reduce to two coupled ordinary differential equations (ODEs),
to which are added
the energy-conservation equation of the matter
and an equation of state $P_{M} = P_{M}\left( \rho_{M} \right)$
[here, we choose a constant equation-of-state parameter
$w_{M}$]. Hence, the cosmological equations are as follows:
\bsubeqs\label{eq:app-model-Friedmann-eqs}
\beqa
\hspace*{-8mm}
\label{eq:app-model-Friedmann-scalar-eq}
&& b^{-2}\,\left(\ddot{\eta} + 3\,\frac{\dot{a}}{a} \, \dot{\eta}\right) =
-\lambda\,b^{-4}\,\big(1-\eta^{2}\big)\,\eta
+\frac{3}{4\pi G}\,\frac{1}{\eta^3}\,
\left[ \frac{\ddot{a}}{a}+ \left( \frac{\dot{a}}{a} \right)^2 \right]\,,
\\[2mm]
\hspace*{-8mm}
\label{eq:app-model-Friedmann-adot-eq}
&&
\left( \frac{\dot{a}}{a} \right)^2
 =
\frac{8\pi}{3}\,G\,\eta^2\,
\left[ \rho_{M} - \frac{1}{2}\,b^{-2}\,\dot{\eta}^2
       + \frac{\lambda}{4}\, b^{-4}\,\big(1-\eta^{2}\big)^2 \right]
- \frac{\dot{a}}{a}\,\eta^2\,\partial_{t}\left[\eta^{-2}\right]\,,
\\[2mm]
\hspace*{-8mm}
\label{eq:app-model-Friedmann-rhoMdot-eq}
&&
\dot{\rho}_{M}+ 3\,\frac{\dot{a}}{a}\,\left( \rho_{M}+ P_{M} \right)=0\,,
\\[2mm]
\hspace*{-8mm}
&&
\label{eq:app-model-Friedmann-w-eq}
\frac{P_{M}(t)}{\rho_{M}(t) } = w_{M}  = \text{const}\,,
\eeqa
\esubeqs
where the last equation gives the equation of state
of the normal matter.

The 11 component of the
gravitational field equation \eqref{eq:app-model-grav-eq} gives
\beqa
\label{eq:app-model-Friedmann-11-grav-eq}
\frac{\ddot{a}}{a} + \frac{1}{2}\,\left( \frac{\dot{a}}{a} \right)^2
&=&
-4\pi G\,\eta^2\,
 \left[ P_{M} - \frac{1}{2}\,b^{-2}\,\dot{\eta}^2
        - \frac{\lambda}{4}\, b^{-4}\,\big(1-\eta^{2}\big)^2 \right]
\nonumber\\[1mm]
&&
+\frac{1}{2}\, \eta^2\,\left[
\frac{\dot{a}}{a}\,\partial_{t}\left(\eta^{-2}\right)
-a^{-3}\,\partial_{t}\left[a^{3}\,\partial_{t}\left(\eta^{-2}\right)\right]\right] \,,
\eeqa
but this equation is redundant.
With a given equation of state \eqref{eq:app-model-Friedmann-w-eq},
the ODE \eqref{eq:app-model-Friedmann-11-grav-eq}
can be shown to follow from the derivative of
the ODE \eqref{eq:app-model-Friedmann-adot-eq},
together with the ODEs \eqref{eq:app-model-Friedmann-scalar-eq},
\eqref{eq:app-model-Friedmann-adot-eq},
and \eqref{eq:app-model-Friedmann-rhoMdot-eq}.
The redundancy of these four ODEs
traces back to the contracted Bianchi identities
and energy-momentum conservation, which, in turn, result from
the general coordinate invariance of the model action
(see App.~E.1 Ref.~\cite{Wald1984}).

At this moment, we observe that the
ODE \eqref{eq:app-model-Friedmann-adot-eq}
is really a \emph{quadratic} in the Hubble parameter
$H \equiv \dot{a}/a$ and we get the following roots:
\beq
\label{eq:app-H-roots}
H_{\pm}
=\frac{\dot{\eta}}{\eta} \pm \sqrt{\left(\frac{\dot{\eta}}{\eta}\right)^2
+\frac{8\pi}{3}\,G\,\eta^2\,
\left[ \rho_{M} - \frac{1}{2}\,b^{-2}\,\dot{\eta}^2
+ \frac{\lambda}{4}\,b^{-4}\, \big(1-\eta^{2}\big)^2 \right]}\,.
\eeq
If the scalar field $\eta(t)$ has the kink-type structure
\eqref{eq:exact-sol-g-zero-eta}, the problem with \eqref{eq:app-H-roots}
is obvious:
$\dot{\eta}/\eta$ diverges at $t=0$ and a bounce-type behavior
with $H(0)=0$ requires, for $t>0$, the minus sign in \eqref{eq:app-H-roots},
which then implies a \emph{contracting} phase for larger values
of $t$, as the $G\,\rho_{M}$ term in the root becomes important.
Hence, removing the big bang at $t=0$ now results in a new
big bang at $t=t_\text{bb}$ for $t_\text{bb}>0$.
For $t<0$, the plus sign in \eqref{eq:app-H-roots} would be
needed. All in all, we do not get a bounce
similar to the one of the upper curve in Fig.~\ref{fig:fig02}.

After we completed the calculations reported in this paper,
we have become aware of earlier papers in the literature,
which discuss bouncing cosmology from a wrong-sign
Brans-Dicke scalar (see, e.g., Ref.~\cite{Tsujikawa2003}
for a research paper and Ref.~\cite{BrandenbergerPeter2016}
for a review of this and other types of models).
Equation (7) of Ref.~\cite{Tsujikawa2003}
has essentially the same quadratic structure
as \eqref{eq:app-model-Friedmann-adot-eq}.
New, here, is the possible role of a kink-type configuration,
which has certain advantages but also spells trouble if there
is a linear term in the $H$ quadratic. For this reason,
we have considered a first-order Friedmann equation of the
form \eqref{eq:model-Friedmann-adot-eq}.\vspace*{0mm}


\section{Possible nonlocal term for the scalar-tensor model}
\label{sec:App-Possible-nonlocal-term}

Inspired by certain results from App.~\ref{sec:App-Brans-Dicke-Model},
we obtain the following \textit{Ansatz} for the symmetric covariant
tensor $X_{\mu\nu}$ appearing
in the  model field equations \eqref{eq:model-field-eqs}:
\beqa\label{eq:app-Xmunu-Ansatz}
\widetilde{X}_{\mu\nu}   &=&
-\frac{2}{3}\;\eta^{-1}\,
\Big[\eta_{,\kappa ;\lambda}- g_{\kappa\lambda}\,\Box\,\eta\Big]\,
\frac{\eta^{,\kappa}\,\eta^{,\lambda}}{(\eta_{,\rho}\,\eta^{,\rho})^2}\,
\Big[\eta_{,\mu}\,\eta_{,\nu}
- g_{\mu\nu}\,\eta_{,\sigma}\,\eta^{,\sigma} \Big]\,.
\eeqa
This term has no direct dependence on the mass scale $1/b$
and is nonlocal due to the presence of the factor
$1/(\eta_{,\rho}\,\eta^{,\rho})^2$.

With the spatially flat Robertson--Walker metric \eqref{eq:RW}
and homogeneous matter fields \eqref{eq:homogeneous-matter-fields},
the \textit{Ansatz} \eqref{eq:app-Xmunu-Ansatz} reproduces the
expressions \eqref{eq:Xmunu-reduced}. In fact, the reduced
form of the last term in square brackets in \eqref{eq:app-Xmunu-Ansatz}
gives precisely the diagonal structure in $\mu\nu$ and a vanishing
00 component. The other terms in \eqref{eq:app-Xmunu-Ansatz}
make that the $mm$ component matches the expression
\eqref{eq:Xmunu-reduced-mm}.
Remark that the limits $a(t) \to 1$ and $\eta(t) \to 1$
of the reduced $\widetilde{X}_{\mu\nu}$ expressions
from \eqref{eq:Xmunu-reduced} define $\widetilde{X}_{\mu\nu}=0$
in the $\eta$-vacuum of Minkowski spacetime.

Turning to the energy-momentum conservation condition
\eqref{eq:X-conservation}, that equation can be simplified
somewhat by use of the scalar field equation \eqref{eq:model-scalar-eq}
and the energy-momentum conservation condition
$T_{\mu\nu}^{(M);\nu}=0$ of the normal matter.
In terms of the contravariant tensor $X^{\mu\nu}$, the simplified
energy-momentum conservation condition reads
\beqa
\label{eq:app-X-conservation-tmp}
&&
X^{\mu\nu}_{\quad;\nu}-16\pi G\,\eta\,\eta_{,\nu}\,
\left[ T^{\mu\nu}_{(\eta)}+T^{\mu\nu}_{(M)} \right]=0\,.
\eeqa
The energy-momentum tensors in the above expression can be
eliminated by use of the model gravitational field equation
\eqref{eq:model-grav-eq}. The result is the following
set of conditions:
\beqa
\label{eq:app-X-conservation}
C^{\mu}
&\equiv&
X^{\mu\nu}_{\quad;\nu}+2\,\left(\eta_{,\nu}/\eta\right)\,
\left[ R^{\mu\nu}- \frac{1}{2}\,g^{\mu\nu}\,R - X^{\mu\nu} \right]=0\,.
\eeqa
These equations effectively correspond to a set of four partial differential
equations for the ten components of the unknown symmetric tensor $X^{\mu\nu}$.

The question, now, is if the
\textit{Ansatz} \eqref{eq:app-Xmunu-Ansatz}
satisfies the condition \eqref{eq:app-X-conservation},
upon use of the field equations \eqref{eq:model-field-eqs}.
We have a few partial answers in the affirmative
(see also App.~\ref{sec:App-Linear-perturbations})
but not a definitive general answer.

\section{Linear perturbations of the kink-bounce solution}
\label{sec:App-Linear-perturbations}

In this appendix, we present some preliminary results
on linear perturbations of the kink-bounce solution as presented
in  Sec.~\ref{subsec:Scalar-kink-solution-with-relativistic-matter}.
For the degenerate-metric bounce, a particular
linear perturbation was found in Ref.~\cite{KlinkhamerWang2020-PRD},
which had a time-independent homogeneous scalar metric perturbation
and a corresponding nonrelativistic-matter perturbation.
Here, we find a similar homogeneous linear perturbation of the
kink-bounce solution. For completeness, we also present results for
an inhomogeneous linear perturbation of the kink-bounce solution.

\subsection{Ans\"{a}tze}
\label{subsec:App-Ansaetze}

The model field equations are given by \eqref{eq:model-field-eqs}
with $\widetilde{X}_{\mu\nu}$ from \eqref{eq:app-Xmunu-Ansatz}
and the energy-momentum tensor of the
normal matter is assumed to be that of
a nonrelativistic perfect fluid,
\beqa\label{eq:app-Linear-perturbations-wM}
w_{M}   &=& 0\,.
\eeqa
We set $b=1$ and use the notation $r_{M}(x)$ for the dimensionless
normal-matter energy density.

Consider, then, the following \textit{Ans\"{a}tze}
for the linear perturbations:
\bsubeqs\label{eq:app-Linear-perturbations}
\beqa
\hspace*{-0mm}
\label{eq:app-Linear-perturbations-metric}
ds^{2}
&=&
- \left[1+2\,\epsilon\,\Phi(t,\,x^1)\right]\, d t^{2}
+ a^{2}(t)\,\left[1-2\,\epsilon\,\Phi(t,\,x^1)\right]
\;\delta_{mn}\,dx^{m}\,dx^n \,,
\\[2mm]
\label{eq:app-Linear-perturbations-eta}
\eta(t,\,x^1)   &=& \chi(t)\, \left[1+\epsilon\,\Xi(t,\,x^1)\right]\,,
\\[2mm]
\label{eq:app-Linear-perturbations-rM}
r_{M}(t,\,x^1) &=& \overline{r}_{M}(t)\,
                   \left[1+\epsilon\,\Sigma(t,\,x^1)\right] \,,
\eeqa
\esubeqs
where the perturbation functions $\Phi(t,\,x^1)$, $\Xi(t,\,x^1)$,
and $\Sigma(t,\,x^1)$ are considered to be of order unity
and $\epsilon$ is a positive infinitesimal.
The boundary conditions on the unperturbed functions
are $a(0)=1$, $\chi(0)=0$,  and $\overline{r}_{M}(0)=r_{M0}\geq 0$.

We will now present two types of solutions for
the perturbation functions $\Phi(t,\,x^1)$, $\Xi(t,\,x^1)$,
and $\Sigma(t,\,x^1)$, the first type being
homogeneous (independent of the spatial coordinate $x^1$) and
the second being inhomogeneous (dependent on $x^1$).

\subsection{Homogeneous scalar metric perturbation}
\label{subsec:App-Homogeneous-scalar-metric-perturbation}

We already have the series solutions
for $\chi(t)$,  $a(t)$, and $\overline{r}_{M}(t)$
from \eqref{eq:pert-sol} and \eqref{eq:pert-sol-coeff} for $w_{M} =0$.
Working up to order $\epsilon$, we then obtain a particular homogeneous
solution of the model field equations \eqref{eq:model-Friedmann-eqs}
around $t=0$,
which has the following structure:
\bsubeqs\label{eq:app-linear-perturbation-series}
\beqa
\Xi(t,\,x^1) &=& c_{00} + c_{20}\,t^2 + \cdots
\,,
\\[2mm]
\Phi(t,\,x^1)&=& d_{00}+ \cdots\,,
\\[2mm]
\Sigma(t,\,x^1)&=& e_{00}+ \cdots\,,
\eeqa
\esubeqs
with coefficients
\bsubeqs\label{eq:app-linear-perturbation-series-coeff}
\beqa
c_{00}&=&
\frac{1}{3 + 27\,\pi\,g }\;
\left(5 + 27\,\pi\,g
- 8\,{\sqrt{3\,\pi\,g\,r_{M0}/\lambda}}\,\right)\,,
\\[2mm]
c_{20}&=& \lambda\,,
\\[2mm]
d_{00}&=&-3\,,
\\[2mm]
e_{00}&=&
\frac{1}{\big(3 + 27\,\pi\,g \big)\,r_{M0}}\;
\Big( \big[7 + 54\,\pi\,g\big]\,\lambda  -
     4\,{\sqrt{3\,\pi\,g\,\lambda\,r_{M0}}}
     + 8\,r_{M0} + 108\,\pi\,g \,r_{M0}
\nonumber\\[1mm]&&
+ 16\,{\sqrt{3\,\pi\,g\,r_{M0}/\lambda}}\;{r_{M0}}\Big)\,,
\eeqa
\esubeqs
where $c_{20}$ has been normalized to $\lambda$
(so that $d_{00}$ is of order unity) and
the dimensionless gravitational coupling constant $g$ has been
defined by \eqref{eq:dimensionless-quantities-g}.
Recall that we have set $b=1$ and that the coordinates $t$ and $x^1$
are dimensionless.

For $g=\lambda=r_{M0}=0^{+}$, the nonvanishing coefficients are
$c_{00}=5/3$, $d_{00}=-3$, and $e_{00}=5$, so that the ratio
$\Sigma(0,\,0)/\Phi(0,\,0)$ equals $-5/3$, which is to be compared
with the ratio $-2$ from the degenerate-metric bounce,
according to Eqs.~(3.19) and (3.22a) in Ref.~\cite{KlinkhamerWang2020-PRD} 
for $\widehat{C}_{\mathbf{k},1} \propto \delta(\mathbf{k})$
and $\widehat{C}_{\mathbf{k},2}=0$.

Turn, now, to the energy-conservation
condition \eqref{eq:app-X-conservation}
with $X_{\mu\nu}=\widetilde{X}_{\mu\nu}$ from \eqref{eq:app-Xmunu-Ansatz},
where the fields from \eqref{eq:app-Linear-perturbations} are inserted,
together with the unperturbed functions
$\chi(t)$,  $a(t)$, and $\overline{r}_{M}(t)$
from \eqref{eq:pert-sol} and \eqref{eq:pert-sol-coeff} for $w_{M} =0$.
Working in orders $\epsilon^0$ and $\epsilon$, the homogeneous
perturbative \textit{Ansatz} \eqref{eq:app-linear-perturbation-series}
then gives certain expressions for $C^{\mu}$
involving the coefficients $c_{00}$, $c_{20}$, $d_{00}$, and $e_{00}$,
with the following behavior near $t=0$:
\beqa
\label{eq:app-Cmu-pert-Ansatze}
C^{\mu}\,\Big|^\text{(hom.\,pert.\,Ansatz)}
&=& \Big( \text{O}(t^5)+\text{O}(\epsilon\,t),\, 0,\, 0,\, 0\Big)^{\mu}\,.
\eeqa
If we, then, use the perturbative solution with
coefficients \eqref{eq:app-linear-perturbation-series-coeff}, we obtain
\beqa
\label{eq:app-Cmu-pert-sol}  
C^{\mu}\,\Big|^\text{(hom.\,pert.\,sol.)}
&=& \Big( \text{O}(t^5)+\text{O}(\epsilon\,t^3),\, 0 ,\, 0,\, 0\Big)^{\mu}\,,
\eeqa
which, near $t=0$, 
is significantly smaller than \eqref{eq:app-Cmu-pert-Ansatze}.

\subsection{Inhomogeneous scalar metric perturbation}
\label{subsec:App-Inhomogeneous-scalar-metric-perturbation}

We, again, have the series solutions
for $\chi(t)$,  $a(t)$, and $\overline{r}_{M}(t)$
from \eqref{eq:pert-sol} and \eqref{eq:pert-sol-coeff} for $w_{M} =0$.
Working up to order $\epsilon$, we now obtain an inhomogeneous
solution of the model field equations \eqref{eq:model-Friedmann-eqs}
around $(t,\,x^1)=(0,\,0)$, which has the following structure 
(with dimensionless coordinates $t$ and $x^1$):
\bsubeqs\label{eq:app-linear-perturbation-series-inhomogeneous}
\beqa
\label{eq:app-linear-perturbation-series-inhomogeneous-Xi}
\Xi(t,\,x^1) &=& c_{01}\,x^1 + c_{21}\,t^2\,x^1 + \cdots\,,
\\[2mm]
\label{eq:app-linear-perturbation-series-inhomogeneous-Phi}
\Phi(t,\,x^1)&=& d_{01}\,x^1 + d_{21}\,t^2\,x^1 + \cdots\,,
\\[2mm]
\label{eq:app-linear-perturbation-series-inhomogeneous-Sigma}
\Sigma(t,\,x^1)&=& e_{01}\,x^1 + e_{21}\,t^2\,x^1 + \cdots\,,
\eeqa
\esubeqs
with coefficients  
\bsubeqs\label{eq:app-linear-perturbation-series-coeff-inhomogeneous}
\beqa        
c_{01}&=& -2\,,
\\[2mm]
c_{21}&=&
-\frac{1}{9}\;
\left(3\, \lambda - 4\, \sqrt{3\, \pi\,g\,\lambda\,r_{M0}}\right)\,,
\\[2mm]
\label{eq:app-linear-perturbation-series-coeff-inhomogeneous-d01}
d_{01}&=&1\,,
\\[2mm]
\label{eq:app-linear-perturbation-series-coeff-inhomogeneous-d21}
d_{21}&=&\sqrt{\pi/3}\; \sqrt{g\,\lambda\,r_{M0}}\,,
\\[2mm]
\label{eq:app-linear-perturbation-series-coeff-inhomogeneous-e01}
e_{01}&=& -\frac{3}{2}\;\lambda/r_{M0}\,,
\\[2mm]
\label{eq:app-linear-perturbation-series-coeff-inhomogeneous-e21}
e_{21}&=&
\frac{1}{18}\;
\left(6\, \lambda+
\sqrt{3\, \pi\,g\,\lambda}\: \big[ 45\, \lambda/\sqrt{r_{M0}}
- 58\, \sqrt{r_{M0}}\,  \big]\,\right)\,,
\eeqa
\esubeqs
where $d_{01}$ has been normalized to unity.  

For $0<g \ll \lambda\sim r_{M0}<1$, the inhomogeneous scalar
metric perturbation $2\,\epsilon\,\Phi$ from
\eqref{eq:app-Linear-perturbations-metric}
and \eqref{eq:app-linear-perturbation-series-inhomogeneous-Phi},
with coefficients \eqref{eq:app-linear-perturbation-series-coeff-inhomogeneous-d01}
and \eqref{eq:app-linear-perturbation-series-coeff-inhomogeneous-d21},
is more or less constant with respect to the cosmic time $t$,
while the relative matter energy density perturbation
$\epsilon\,\Sigma$ from
\eqref{eq:app-Linear-perturbations-rM}
and \eqref{eq:app-linear-perturbation-series-inhomogeneous-Sigma},
with coefficients \eqref{eq:app-linear-perturbation-series-coeff-inhomogeneous-e01}
and \eqref{eq:app-linear-perturbation-series-coeff-inhomogeneous-e21},
ultimately grows with cosmic time $t$.
This behavior is reminiscent of the one found 
for the degenerate metric bounce, as given by Eqs.~(3.17) and (3.22b) 
in Ref.~\cite{KlinkhamerWang2020-PRD} for
$C_1(\mathbf{x})=x^1$ and $C_2(\mathbf{x})=0$ or
$\widehat{C}_{\mathbf{k},2}=0$.


Turn, again, to the energy-conservation
condition \eqref{eq:app-X-conservation}
with $X_{\mu\nu}=\widetilde{X}_{\mu\nu}$ from \eqref{eq:app-Xmunu-Ansatz},
where the fields from \eqref{eq:app-Linear-perturbations} are inserted,
together with the unperturbed functions
$\chi(t)$,  $a(t)$, and $\overline{r}_{M}(t)$
from \eqref{eq:pert-sol} and \eqref{eq:pert-sol-coeff} for $w_{M} =0$.
Working in orders $\epsilon^0$ and $\epsilon$, the inhomogeneous
perturbative \textit{Ansatz}
\eqref{eq:app-linear-perturbation-series-inhomogeneous}
then gives somewhat complicated expressions for $C^{\mu}$
involving the coefficients
$c_{01}$, $c_{21}$, $d_{01}$, $d_{21}$, $e_{01}$, and $e_{21}$,
with the following behavior near $(t,\,x^1)=(0,\,0)$:
\beqa
\label{eq:app-Cmu-pert-Ansatze-inhomogeneous}
C^{\mu}\,\Big|^\text{(inhom.\,pert.\,Ansatz)}
&=& \Big( \text{O}(t^5)+\text{O}(\epsilon\,t\,x^1),\, \text{O}(\epsilon),\, 0,\, 0\Big)^{\mu}\,,
\eeqa
where, generically, the second component is nonvanishing
for small but finite $\epsilon$.
If we, then, use the perturbative solution with coefficients
\eqref{eq:app-linear-perturbation-series-coeff-inhomogeneous}, we obtain
\beqa
\label{eq:app-Cmu-pert-sol-inhomogeneous}  
C^{\mu}\,\Big|^\text{(inhom.\,pert.\,sol.)}
&=& \Big( \text{O}(t^5)+\text{O}(\epsilon\,t^3\,x^1),\, 
\text{O}(\epsilon\,t^2) ,\, 0,\, 0\Big)^{\mu}\,,
\eeqa
which, near $(t,\,x^1)=(0,\,0)$,
is significantly smaller than \eqref{eq:app-Cmu-pert-Ansatze-inhomogeneous}.

\end{appendix}

\newpage

\end{document}